# Ab initio calculation of endohedral fullerenes with various metal atoms

V.S. Gurin

*Physico-Chemical Research Institute,*
*Belarusian State University, Leningradskaya str., 14, Minsk, 220080, Belarus;*
*E-mail: gurin@bsu.by; gurinvs@lycos.com*

Abstract

The model of endofullerenes $M@C_{60}$ with M=Li,Na,Cs,Cu,Ag are proposed for study an effect of the nature of endoatoms upon features of the minimum energy structures. They were calculated by SCF Hartree-Fock method with ECP (Ag and Cs) and all-electronic basis sets assuming arbitrary symmetry distortion ($C_1$ point group). All models display the off-center position for the endoatoms and the charge transfer between M and the carbon cage strongly dependent on the nature of endoatoms. Ionization potentials and atomic radii are proposed as the basic factors providing the properties of endofullerenes.

*Keywords*: fullerene, Hartree-Fock calculations, metal atoms, asymmetry

## 1. Introduction

Endohedral fullerenes $M@C_{60}$ are of great interest from the point of both experimental and theoretical studies of new type of chemical species [1-3]. Their formation occurs due to the geometrical confinement of atoms inside the closed carbon cage $C_{60}$. A question on chemical bonding of an endoatom with carbon *a priori* is open since the simple geometrical reasons provide the possibility of these systems. However, bare metal atoms as well non-metal elements in atomic form usually are strongly reactive, and the M-C bond formation is very probable. It is known that non-metal atoms such as nitrogen, phosphorus can produce the endohedral structures M@$C_{60}$ with negligible effect upon the carbon cage [4,5], while the active metals (alkali, rare earth elements) interact much stronger with $C_{60}$ [2,3,5,6]. Weakly interacting atoms are localized in the center of $C_{60}$ molecule, and the electronic state of these atoms deserves a special study [7]. The state of metal atoms inside $C_{60}$ has been calculated in many works (see review in Refs. 2,3, however, the symmetrical structures were considered in more degree as far as an initial $C_{60}$ possesses the high icosahedral symmetry. In the present work, we calculate a series of M@$C_{60}$ models assuming arbitrary symmetry distortion to demonstrate how the nature of endoatoms influences the properties of the endofullerenes. M@$C_{60}$ with M=Li, Na, Cs are really existing species [8-10] while the species with M=Ag and Cu are not well established yet [11,12], however, there are arguments on possibility of their existence [13].

## 2. Calculation method

For calculations we used the *ab initio* SCF Hartree-Fock method with accounting electron correlation effects for selected structures at the MP2 level. Basis sets were: effective core potential (ECP) with 28-e core for Ag, 46-e core for Cs, and all-electronic basis sets for Cu of 6-31G* quality, and STO-3G and 6-31G for C and Na, respectively. The geometry optimization of the models was done at the two levels: (i) with initial symmetry conservation ($I_h$) placing endoatoms into the center and (ii) under arbitrary distortion down to $C_1$ point group that allowed any off-center shift of endoatoms. NWChem software (versions 4.3-4.5) [14] was utilized for the calculations.

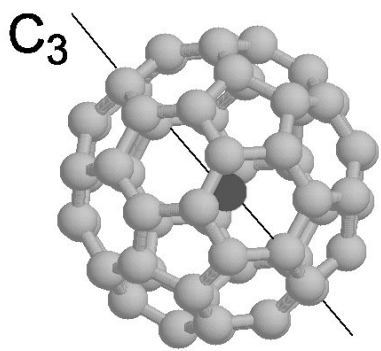

**Fig. 1.** A view of M@$C_{60}$ endofullerene with the $C_3$ axis shown.

## 3. Results and Discussion

The calculation results for geometry of the models, binding energy and effective charges at endoatoms derived from the Mulliken occupancy analysis are summarized in Table 1. Fig. 1 displays the general view of optimized structures in which an endoatom is shifted off the center. $C_3$ axis is one of the symmetry axes of $C_{60}$ molecule ($C_2$ and $C_5$ are the other two ones), and it appeared to be the important parameter in the geometry of the optimized models of M@$C_{60}$ under study.

Table 1. Calculated characteristics of the endofullerenes with a series of metal atoms

| Model | Minimum C-C distance Å | Maximum C-C distance Å | Off-centre position of the metal atoms Å | Binding energy of M-$C_{60}$ system, eV | Net charge of M |
|---|---|---|---|---|---|
| $C_{60}$ | 1.38 | 1.46 | | | |
| Ag@$C_{60}$ | 1.40 | 1.54 | 0.28 | 10.2 | -0.10 |
| Cu@$C_{60}$ | 1.41 | 1.54 | 0.48 | 2.8 | +0.35 |
| Li@$C_{60}$ | 1.41 | 1.54 | 1.10 | 7.6 | -0.15 |
| Na@$C_{60}$ | 1.41 | 1.54 | 1.30 | 13.2 | +0.85 |
| Cs@$C_{60}$ | 1.41 | 1.54 | 0.07 | 19.8 | +1.19 |

The shift of endoatoms according to the calculations occurs approximately along the $C_3$ axis. It is of interest that the calculations keeping $C_3$ symmetry (i.e. the placement of the endoatoms along this axis) resulted no in the lower energy. Thus, the most stable structures are asymmetrical.

The value of the off-center shift depends strongly on the nature of metal atom inside. maximum shift is for alkali atoms besides Cs that is different from Li and Na by its atomic radius. Effective charges are rather featured not only in their values but also in the signes. The big positive charges for Na and Cs are understandable as familiar property of endoatoms to be an electronic donor, however, the inversed sign in the case of Li is not expectable from this point of view. The anomalous charge is observed also for silver. Ag is not efficient electronic donor and possesses a large atomic radius. As for binding energies (Table 1) we note that all these models correspond to possible stable structures, while the rather large binding energy for Ag@$C_{60}$ is of interest, though, it cannot be compared with experiment yet.

As possible interpretation of these results we suggest the following three factors responsible for features of the endofullerenes:

(i) Ionization potentials for a series of the metal vary in the sequence Ag > Cu > Li > Na > Cs that results in the corresponding sequence of effective charges from negative one to the superionization, besides Li which is anomalous due to very small radius and large mobility.

(ii) Atomic (ionic) radii for the metals are in the following relationship: Ag > Cu and Cs > Na > Li. Thus, the larger atoms are shifted from the center of the cage in less degree.

(iii) The contribution of different atomic orbitals into the metal-$C_{60}$ bonding: in the case of Ag and Cu *ns* and *(n-1)d* are active while for Li and Na only *ns*-AO, but for Cs *5p* AO become to be of importance. Na behaves as the typical *s*-metal, and Li is also similar, but Li is too small and aspires to penetrate the carbon cage providing strong interaction with it and anomalous charge. The anomalous value of effective charge in the case of Ag can be explained by participation of d-AO (confirmed by the occupancy analysis).

## 4. Conclusions

A series of endofullerenes M@$C_{60}$ with M=Li,Na,Cs,Cu,Ag were calculated at the ab initio level assuming arbitrary symmetry distortion. The off-center position of the endoatoms strongly dependent on the nature of them is shown for the minimum energy structures together with anomalous charge transfer for Li@$C_{60}$ and Ag@$C_{60}$. The variation of the properties in the sequence of endofullerenes are interpreted as contribution of three basic factors: ionization potential of endoatoms, their atomic radius, and participation of d-orbitals in the M-C bond formation.

**Acknowledgements**

The authors want to acknowledge the support of this work due to participation in the project under the Ministry of Education of Belarus.